\documentclass[aps,prl,preprint,showpacs,twocolumn,preprintnumbers,amsmath,amssymb,superscriptaddress,footinbib]{revtex4}
\usepackage{graphicx}
\usepackage{dcolumn}
\usepackage{bm}
\begin{document}

\title{Tricritical behavior in MnSi at nearly hydrostatic pressure}

\author{Alla E. Petrova}
\affiliation{Institute for High Pressure Physics of Russian
Academy of Sciences, Troitsk, Moscow Region, Russia}
\author{Vladimir Krasnorussky}
\affiliation{Institute for High Pressure Physics of Russian
Academy of Sciences, Troitsk, Moscow Region, Russia}
\author{John Sarrao}
\affiliation{Los Alamos National Laboratory, Los Alamos, NM,
87545}
\author{Sergei M. Stishov}
 \email{sergei@hppi.troitsk.ru}
\affiliation{Institute for High Pressure Physics of Russian
Academy of Sciences, Troitsk, Moscow Region, Russia}
\affiliation{Los Alamos National Laboratory, Los Alamos, NM,
87545}

\date{\today}

\begin{abstract}
AC magnetic susceptibility of a single crystal of MnSi was
measured along the ferromagnetic phase-transition line up to a
pressure of 0.8 GPa created by compressed helium. The results show
that the tricritical point is situated at much lower pressure and
at significantly higher temperature ($P_{tr}\cong0.355$ GPa,
$T_{tr}\cong25.2$ K) than was reported previously ($\sim1.2$ GPa,
$\sim12$ K). These new observations put certain constraints on the
origin of the tricritical point in MnSi.
\end{abstract}

\pacs{62.50.+p, 64.60.Kw, 75.30.Kz}

\maketitle Recently, there has been a considerable interest in
studying the magnetic phase transition in the intermetallic
compound MnSi \cite{1,2,3,4,5,6,7,8,9,10,11}, which reveals weak
ferromagnetic properties slightly below 30 K with spins having
been ordered into a long wavelength helical structure
\cite{12,13}. As shown for the first time in \cite{1}, the Curie
temperature $T_{c}$ in MnSi decreases on compression and is
completely suppressed at pressure more than 1.5 GPa. One of the
important findings of earlier studies was the discovery that the
magnetic transition in MnSi becomes first order at pressures
greater than 1.2 GPa, with a tricritical point at $\sim1.2$ GPa,
$\sim12$ K \cite{2,4,7}.  A number of ideas were suggested to
explain this feature of the transition line in MnSi \cite{norm},
including the existence of a deep minimum in the density of state
at the Fermi level \cite{4}, coupling of the order parameter and
soft particle-hole excitations \cite{15}, magnetic rotons
\cite{16}, etc. On the other hand, nonhydrostatic components of
pressure also could change or influence the nature of the phase
transition. In all previous experimental studies, various kinds of
liquids, which freeze to strong solids on cooling, were used as
pressure media. Consequently, it is appropriate to carry out a
study of the magnetic phase transition in MnSi at high pressures
making use of a hydrostatic pressure-transmitting medium.

\begin{figure}[htb]
\includegraphics[width=35mm, height=80mm]{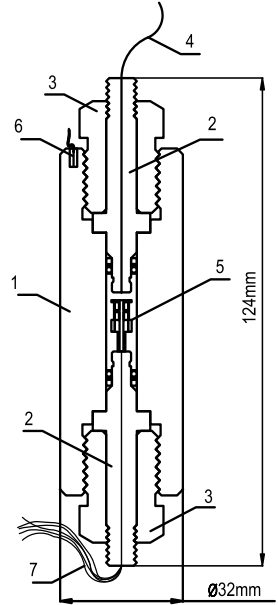}
\caption{\label{fig1} Schematic drawing of high pressure cell: 1)
cell body, 2) plugs, 3) holding nuts, 4) high-pressure tubing, 5)
coil setup, 6) temperature sensor, 7) electrical leads. }
\end{figure}

In the present communication, we report results of our studies of
the AC magnetic susceptibility of a single crystal of MnSi at high
pressure using fluid and solid helium as a pressure medium. Solid
helium, being a quantum crystal, exhibits the highest plastic
properties and is the best medium to transmit pressure at low
temperatures. Single crystals of MnSi were grown from a tin flux
by dissolving pre-alloyed Mn and Si in excess Sn. The resulting
crystals were characterized by powder x-ray diffraction,
electrical resistivity, and magnetic susceptibility. The
temperature of the magnetic phase transition was $29.1\pm0.02$ K.
The resistivity ratio $R_{300}/R_{(T=4.2)}$ varied in the range 80
-100 for different crystals. Descriptions of the high pressure gas
installation and the cryostat are given in \cite{17}. Briefly, the
experimental cell Fig.\ref{fig1} is connected to the high pressure
generator by means of a stainless steel high pressure tubing of
1.6 mm in outer diameter. Details of the experimental cell, made
of beryllium copper, are seen in the Fig.1. Fifteen copper wires
were planted inside the cell using 2850FT Stycast epoxy as a seal
and an insulator. Special measures were taken to make this seal
gas tight at high pressure and low temperature \cite{18}.

To measure AC susceptibility a three-coil set up (drive coil and a
balanced pair of pick up coils) was mounted inside the high
pressure cell. The measurements were carried out with a standard
modulation technique at a modulation frequency 19 Hz.

\begin{figure}[htb]
\includegraphics[width=80mm, height=65mm]{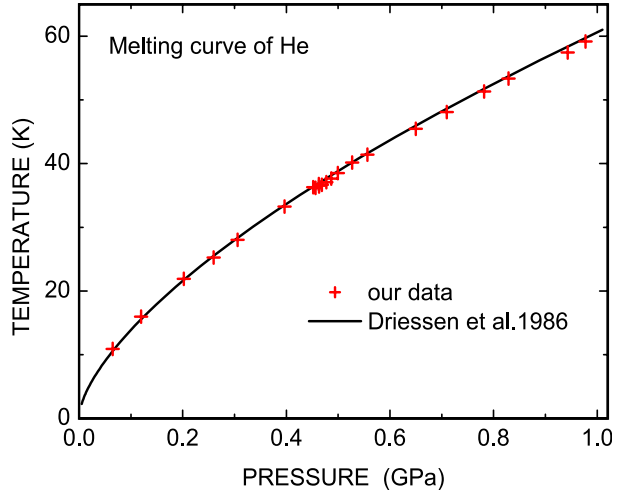}
\caption{\label{fig2} Experimentally determined melting curve of
helium. Solid line - generalized melting curves of helium, taking
into account many fine measurements (exhausting references see in
\cite{19}).
 }
\end{figure}

Temperature was measured by a calibrated Cernox sensor imbedded in
the cell body as shown in the Fig.\ref{fig1} (typical accuracy of
the Cernox sensor in the temperature range under study is about
0.02 K). A calibrated manganin gauge, situated in a separate
temperature stabilized cell, was used to measure pressure with
accuracy about 0.001 GPa while helium was still in the fluid
phase. Upon crystallization of helium, pressure was calculated on
the basis of the measured helium-crystallization temperatures and
extensive data on EOS of helium \cite{19}. Normally the
measurements were taken in the regime of slow cooling of about
-0.2 K/min. On cooling a slight drop of pressure signaled helium
crystallization. A comparison of the measured melting temperatures
of helium with data taken from literature shows that our data do
not deviate much from the precise data obtained in different
laboratories Fig.\ref{fig2}. This confirms that there was no
significant temperature gradient in our pressure cell. Note that
the deviations, being random at low pressures, acquire a more
systematic character when pressure increased, which probably
indicates pressure calibration problems. As a whole, overall
accuracy of the melting temperature of helium, determined in our
experiments, can be estimated as 0.2 K, which puts a lower limit
of about 0.005 GPa for the error in pressure estimations in solid
helium.

\begin{figure}[htb]
\includegraphics[width=80mm, height=65mm]{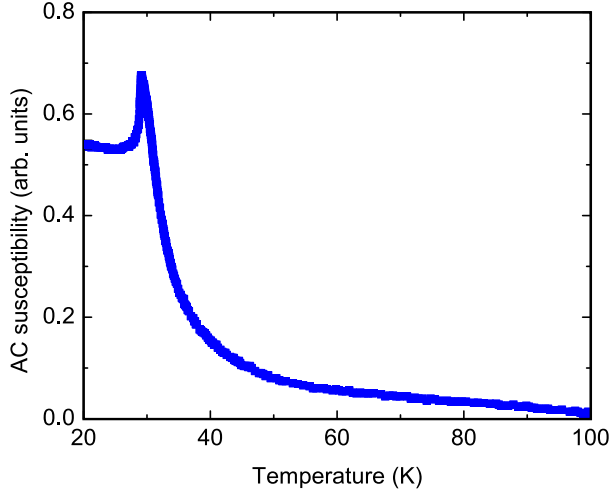}
\caption{\label{fig3} Temperature dependence of AC magnetic
susceptibility $(\chi_{AC})$ of MnSi at ambient pressure. }
\end{figure}

\begin{figure}[htb]
\includegraphics[width=80mm, height=65mm]{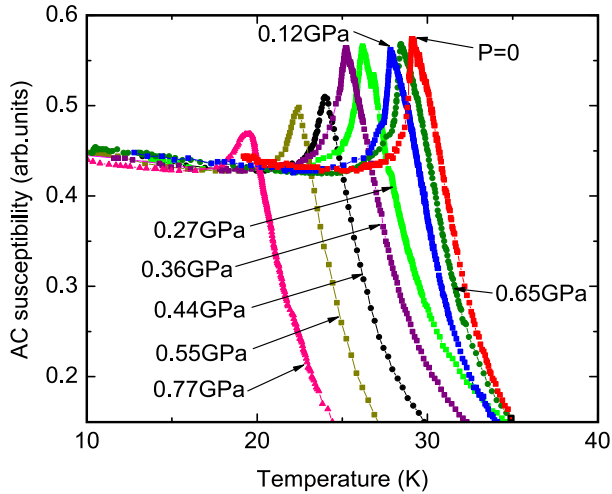}
\caption{\label{fig4} Influence of pressure on the $\chi_{AC}$
singularity at the phase transition in MnSi. }
\end{figure}

Behavior of the AC magnetic susceptibility $(\chi_{AC})$ of MnSi
in a wide temperature range and at ambient pressure is shown in
Fig.\ref{fig3}. Selected curves in Fig.\ref{fig4} demonstrate the
influence of pressure on the position and form of the $\chi_{AC}$
singularity at the magnetic transition in MnSi.


\begin{figure}[htb]
\includegraphics[width=80mm, height=65mm]{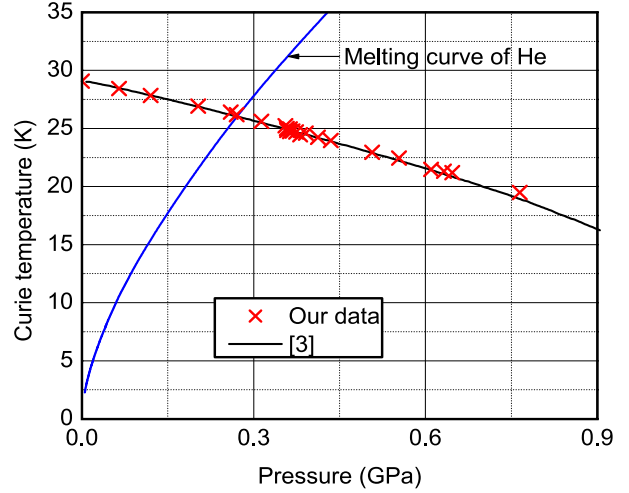}
\caption{\label{fig5} Pressure dependence of the Curie temperature
in MnSi determined in the current experiment. }
\end{figure}

\begin{figure}[htb]
\includegraphics[width=80mm, height=65mm]{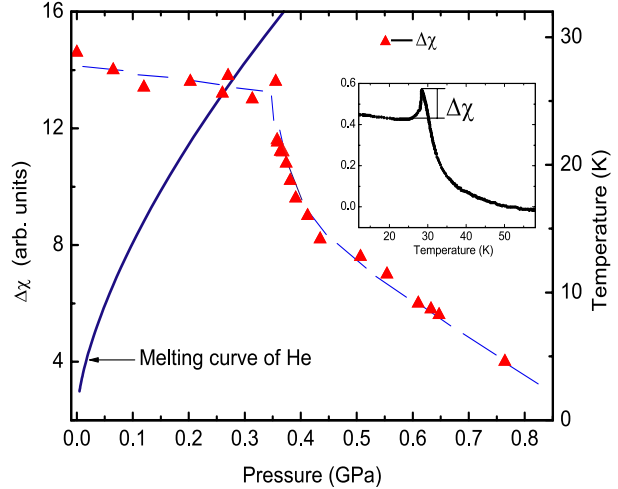}
\caption{\label{fig6} Evolution of the form of the singularity of
$\chi_{AC}$ at the phase transition in MnSi. (Temperature scale on
the right ordinate is related to the melting curve of He.) The
form of the singularity is characterized by the quantity
$\Delta\chi$, illustrated in the insert.
 }
\end{figure}

Temperatures corresponding to the maximum of $\chi_{AC}$ were
taken as the phase-transition temperatures. Consequently, accuracy
of this procedure depended on sharpness of the corresponding
maxima, the experimental resolution and the noise, and varied from
0.02 K to 0.5 K with the average accuracy of about 0.1 K. This
number leads to an average pressure uncertainty of the
phase-transition line of $\sim0.009$ GPa, which probably absorbs
the errors connected with the pressure estimations in solid
helium. The phase-transition line of MnSi in the $P-T$ plane is
depicted in the Fig.\ref{fig5}. As seen from Fig.\ref{fig5}, our
results agree well with the data obtained previously \cite{2,3}.
This comparison shows that effects of nonhydrostatic environments
do not influence much the position of the phase-transition line,
at least in the pressure range probed in the current research, but
as we will see next, nonhydrostaticity does influence the form of
the $\chi_{AC}$ curve near the phase transition and, consequently,
influences the conclusions that were made thereafter.
As argued earlier \cite{2,3}, the temperature dependence of
$\chi_{AC}$ near the phase-transition boundary reflects the order
of the transition. According to \cite{2,3} a $\lambda$-type
singularity of $\chi_{AC}$, typical of the second order phase
transitions, continuously deforms on increasing pressure to become
finally (at $P\cong1.2$ GPa) a simple step expected for first
order transitions. Under the hydrostatic conditions used in the
present experiments, a $\lambda$-type singularity of $\chi_{AC}$
at the phase transition in MnSi remains essentially independent on
pressure to 0.355 GPa, even though some of the data points are
well inside the solid helium field Fig.\ref{fig6}. Just beyond
0.355 GPa, however, the peak in $\chi_{AC}$ starts to decrease
drastically. This evolution in the pressure-dependent shape
$\chi_{AC}(T)$ is just what one would anticipate from the physics
of tricritical phenomena. Our current conclusion, then, is that
the tricritical point on the phase-transition line is situated at
much lower pressure and significantly higher temperature
($P_{tr}\cong0.355$ GPa, $T_{tr}\cong25.2$ K) than was reported in
previous work \cite{2,3} ($\sim1.2$ GPa, $\sim12$ K). The reason
for the observed difference is most probably connected with random
strains caused by nonhydrostatic compression. In that case, the
random strains would play the role of impurities, smearing out
singularities in physical quantities at the second order phase
transition. Finally, the current observation probably does not
support theories of the trictitical phenomena in MnSi that exploit
the specifics of electron spectra of MnSi or metals in general if
that theory is relevant just at zero temperature. The volume
change associated with a first order phase transition clearly
implicates lattice involvement in the mechanism of the tricritical
behavior in MnSi, and the spin-lattice interaction should be taken
into consideration in future theories. Studies of phonon spectra,
thermal expansion and the compressibility of MnSi at high
pressure, created by compressed helium, seem to be indispensable
for a complete explanation of the $P-T$ phase diagram of MnSi.
\begin{acknowledgments}
In conclusion, the authors express their gratitude to Dr. Joe
Thompson for reading the manuscript and valuable suggestions.\\
A.E. Petrova, V. Krasnorussky and S.M. Stishov appreciate support
of the Russian fund for Basic Research (grant N03-02-17119),
Program of the Physics Department of Russian Academy of Science on
Strongly Correlated Systems and Program of the Presidium of
Russian Academy of Science on Physics of Strongly Compressed
Matter.
\end{acknowledgments}

\bibliography{basename of .bib file}

\begin{thebibliography}{99}
\bibitem{norm} Normally magnetic ordering in the itinerant systems occurs as a
second order phase transitions that can be well described in the
framework of a mean-field theory.  Emerging of a tricritical point
then means that the forth order term of the Landau expansion
change the sign \cite{14}. The latter takes place in result of
coupling of an order parameter with others not critical degree of
freedom.
\end{thebibliography}

\end{document}